\documentclass[prb,preprint]{revtex4-1} 

\usepackage{amsmath}  
\usepackage{amsfonts} 
\usepackage{graphicx} 

\begin{document}

\title{A Colorful Demonstration of Thermal Refraction}

\author{Mark Ciotola}
\email{ciotola@sfsu.edu} 
\altaffiliation[ ]{1600 Holloway Avenue FA121, San Francisco, CA 94132} 
\affiliation{Department of Design and Industry, San Francisco State University, San Francisco, CA 94132}

\author{Olivia Mah}
\email{ommah@usfca.edu}
\affiliation{Department of Mathematics, University of San Francisco, San Francisco, CA 94117}

\date{\today}

\begin{abstract}
The wave nature of heat flow mechanisms, such as lattice waves is discussed. Tan and Holland's Tangent Law of heat flow refraction is reviewed. A classroom demonstration of heat flow refraction through conductors in series is presented, and sample results are examined for consistency with the Tangent Law. To predict results, the Tangent Law is derived from the Principle of Least Resistance for this demonstration. User-modifiable simulations in Ruby and Ruby on Rails are presented, along with simulation results for various combinations of conductors. Results are interpreted in terms of the Principle of Least Time, illustrating a powerful unification in physics between disparate areas such as optics and thermodynamics.
\end{abstract}

\maketitle 

\section{Introduction} 

The use of  the Principle of Least Time (Fermat's Principle) to derive Snell's Law of Refraction\cite{Brizard} is well known and commonly presented in introductory physics textbooks. Yet while it is common knowledge that light rays are refracted, less commonly known is that heat flow is also refracted. 

Such is unsurprising when one considers that heat is propagated by particles of matter that, like photons, exhibit a wave-particle duality. In solids, ``carriers of energy most frequently encountered are the lattice waves and the free electrons.''\cite{Klemens1969} It is well known that electrons have a DeBroglie wavelength of $\lambda = h/p$.\cite{Resnick1985} Heat flow in a metal chiefly takes place due to the movement of free electrons\cite{gsu_2} that can be viewed as a highly degenerate electron gas.\cite{Klemens1969} So as electrons transport thermal energy across a series of metals of differing thermal conductivities, it is reasonable to expect the wave nature of those electrons to experience refraction. 

The refraction of heat flow is supported in the literature. Holland discusses an experiment concerning heat flow from a crucible into both solid and liquid silicon, and between each other. Heat flow is observed to be refracted and an analog of Snell's Law is described. \cite{Holland1989} Tan and Holland disclose a tangent law (hereinafter the Tangent Law) for thermal refraction of heat flow across a boundary of materials with differing thermal conductivities. The Tangent Law is analogous to Snell's Law, but must be adjusted due to the constraint of heat flow to mathematical tubes, since the heat flow is planer: \cite{Tan1990} 
\begin{equation}\label{E:tangent_law_w_k}
 \frac{tan{\theta_1}}{k_1} = \frac{tan{\theta_2}}{k_2}, 
\end{equation}
where $k_1$ is the thermal conductivity of conductor 1 and  $k_2$ that of conductor 2.

Further, Bertolotti et al. shows experimental evidence for how a harmonically varying heat source produces thermal waves that are refracted when approaching the interface between two media.\cite{Bertolotti1999} Shendeleva discloses reflection and refraction of a plane thermal wave of oblique incidence at an interface.\cite{Shendeleva2003} Burt describes combinations of metal that form thermal lenses that either focus or spread heat rays.\cite{Burt2006}

\section{A Colorful Classroom Experiment}

According to Fourier's Law of Heat Conduction, the heat energy flow through a conductor bridging a thermal difference is proportional to the conductor's area $A$ and the thermal difference bridged by temperature difference $\Delta T$, and is inversely proportional to the conductor's length $L$.\cite{Schroeder2000} This law was formulated by Joseph Fourier in 1822, \cite{Davies} and can be stated as:

\[ \frac{dQ}{dt} = k \Delta T \frac{A}{L}, \]
where $Q$ is heat energy, $t$ is time, $k$ is the conductor material's thermal conductivity.

However, Fourier's Law only describes heat flow in a simple linear path through a single medium.  Describing heat flow through multiple media can be achieved by the Principle of Minimum Thermal Resistance.\cite{Holland1989} If the greatest proportion of heat flows through the path of least thermal resistance, then the total rate of heat flow is maximized. (Feynman et al. provides a more illustrative discussion of how the Principle of Least Time operates over multiple paths. \cite{Feynman1964})

A series  of two media of different thermal conductivities is the simplest to model. The path of minimum thermal resistance will take a longer route through the material with greater thermal conductivity and a shorter route through the material with lesser conductivity. Hence, the flow of heat is refracted. This is similar to the refraction of light in Snell's Law, where the proportional velocity of photons through each medium determines the light's path. When the most heat flows through the path of minimum thermal resistance, then a particular quantity of heat shall flow through the media series in the least amount of time. Thus, both the refraction of light and heat flow are a consequence of the Principle of Least Time. 

A blackbody, such as a tungsten filament, will attempt to emit both heat and light in all directions. A light ray can be created by blocking the blackbody with a mask permitting light to only pass through a small hole or line. It is likewise conceivable to create a heat ray by similarly masking a source of heat so that most of the heat energy flows in nearly the same direction. In the case of conduction electrons, the ray will travel substantially intact through several mean free paths. However, we do not need to resort to such microscopic considerations to demonstrate thermal refraction. All that is required is a heat source and sink. The somewhat nebulous flow of heat is handled by the Tangent Law.

\begin{figure}[h!]
\centering
\includegraphics{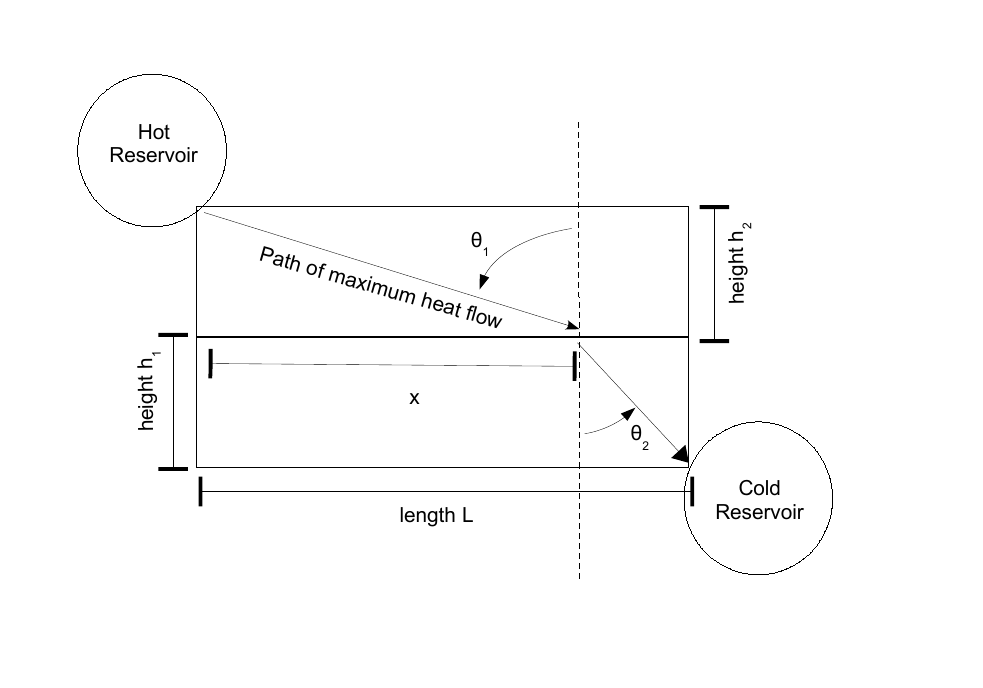}
\caption{Heat flowing through two conductors in parallel.}
\label{thermal_refraction_01}
\end{figure}

The authors have set up an experiment involving heat flow through two materials in series with different thermal conductivities. \cite{Ciotola2003} The materials are inexpensive bars of metal that are 5.0 cm  long and 1.3 cm wide and high that are available through science education supply firms. The bars are arranged parallel lengthwise and firmly touch each other along one side of their length. Heat is allowed to flow into one extreme corner of one conductor and exit through the opposite corner of the other conductor. (See Fig.~\ref{thermal_refraction_01}.) This is managed by placing a hot object at one corner and a cold object at the other corner. Temperature sensitive indicator film is placed on top of the conductors. Except for these corners, the conductors should be completely surrounded by a good insulator, both beneath and at the sides of the conductors, otherwise valid results will likely not be obtained. Styrofoam is a good insulator for this demonstration in that it is easily shaped, freely available from waste materials, and not damaged by water. 

The surface between the two bars should be wetted with thermally conductive paste or water, to ensure a good thermal contact (pressure against the lengthwise dimension of the bars further improves contact). A small metal cup, such a paint cup for an artist's palette, filled with hot water suffices for the hot object (take care to avoid burns). A second metal cup filled with ice water, or even just a small piece of ice, suffices for the cold object. 

Temperature sensitive liquid crystal display (LCD) film can be used to indicate variations of temperature. 20--25$^\circ$C range film works best. It can be cut to just cover the two metal bars, and should be coated with thermal conduction paste or water on the underside for improved thermal contact, but kept dry on the visible side to avoid cooling due to evaporation. Film areas with the same color have the same temperature and hence represent isotherms. Heat flows perpendicularly to isotherms, or very nearly so. The progression of isotherms indicate temperature gradients.

For this experiment, there are two chief scenarios. Either the conductivities of the media are similar or dissimilar. Where they are rather dissimilar, the more conductive medium acts literally as an express lane for heat flow. For media with similar conductivities, refraction is less pronounced.

Conducting the demonstration to qualitatively verify whether the results are consistent with the Principle of Least Resistance is simple. Have the students note in which conductor the isotherms are the most perpendicular to the long dimension (length) of the conductor. If this is the material with the greatest conductivity, then the Principle is confirmed, since the chief path of heat flow is taking a longer path through the more conductive medium and resistance is thus minimized. Conversely, the conductor with isotherms that are most parallel to the long dimension should be the material with least conductivity. 

For a quantitative experiment, students should record the angle of the most distinctive isotherm in each conductor. This may require patience and several trials, and careful attention to color variation (taking photograph is helpful, but optional). Pressing an ice cube against the conductor corner speeds up the demonstration, but should be sparingly applied. Relaxation time may be required to overcome the effect of the boundary between conductors. Good lighting is helpful, but lamps should not provide so much heat that the results are distorted. The uncertainty should also be noted. It is visually reasonable to distinguish between 15, 30, 45, 60, 75 degree angles, each which differs by 15 degrees, so a baseline uncertainty of 15 degrees is reasonable for this set-up. Half of 15 degrees is possible in principle, but only if sufficiently great care is taken, which may be beyond the limits of a simple ``home-made'' classroom  demonstration. Further improvement may be possible with better insulation, electric heat and cold sources (or a laser for the heat) and digital imaging.

The standard school metal bar set includes one bar each of aluminum, brass, copper and iron. The combination of copper and iron provides the greatest thermal conductivity difference and hence the most dramatic result. If two sets of bars are obtained, then control cases can be set up by pairing together the copper bars, then the iron bars, etc. No refraction should be observed in the control cases, except to the extent that the boundary effect between conductors has not been minimized.

Experimental results for various pairs with uncertainties are shown in Table~\ref{experimentresults}. The angles shown are for the path of maximum heat flow, which is simply the isotherm angle subtracted from 90$^\circ$. The first member of the pair is closest to the heat source, while the second is closest to the heat sink. Results may vary due to the placement of insulation and hot and cold sources and unmitigated boundary effects.

\begin{table}[h!]
\centering
\caption{Experimental Results for Copper and Iron}
\begin{ruledtabular}
\begin{tabular}{l c c c c}
Pair of materials & Observed Angle $\theta_1$  & Uncertainty & Observed Angle $\theta_2$ & Uncertainty\\
\hline
Copper--Iron & 75$^{\circ}$ &  +/- 15$^{\circ}$ & 45$^{\circ}$ & +/- 15$^{\circ}$ \\
Iron--Copper & 30$^{\circ}$ &  +/- 15$^{\circ}$  & 85$^{\circ}$ & +/- 15$^{\circ}$ \\
Copper--Copper & 60$^{\circ}$ &  +/- 15$^{\circ}$ & 75$^{\circ}$ & +/- 15$^{\circ}$\\
Iron--Iron & 60$^{\circ}$ &  +/- 15$^{\circ}$ & 75$^{\circ}$ & +/- 15$^{\circ}$ \\
\end{tabular}
\end{ruledtabular}
\label{experimentresults}
\end{table}

As mentioned, Copper--Iron gives the most dramatic result due to their greatly different thermal conductivities. The copper conductor is clearly dominant in terms of length of maximum flow path. Briefly pressing and melting an ice cube against the corner of the iron conductor gives it a bit more ``fight'', so one can more dynamically examine the interplay in heat flow among the two conductors. Iron--Iron provides the best visual range of colors due the the relative slowness of the thermal flow. There is an observed asymmetry in the copper--iron versus iron--copper pairings. This may be due to a thermal barrier formed at the boundary of the two conductors or where the heat sink has more effect than the heat source. This asymmetry will vary due to experimental set-up, and would need to be taken into account to reduce error.

\section{Expected Results}

To quantitatively verify whether the results are consistent with the Principle of Least Resistance, plug Angle 1 into the Tangent Law:

\[	  \theta_2 =   \arctan{\left(\frac{k_2}{k_1} \tan{\theta_1}\right)}, \].

Use the uncertainties to calculate the upper and lower limits of what is expected. Note that total uncertainty must be considered, since two measurements are involved. This involves simply adding the individual uncertainties. In the set-up discussed, this is $(\pm 15^{\circ}) + (\pm 15^{\circ})$, for a range either way of up to $30^{\circ}$ (we will do better in the next section). This range, though large, is sufficient to reject several hypothetical cases which are inconsistent with heat refraction, or to detect an ineffective experimental set-up, such as due to insufficient insulation.

Experimental results for Angle 2 as a function of Angle 1 are tested for consistency with the Tangent Law for various pairs of materials in Table~\ref{resultsvtangentlaw}.

\begin{table}[h!]
\centering
\caption{Observed versus expected Angle $\theta_2$ where Angle $\theta_1$ is known}
\begin{ruledtabular}
\begin{tabular}{l c c c}
Pair of materials & Observed Angle $\theta_2$ & Expected Angle  $\theta_2$ & Total uncertainty (range) \\
\hline
Copper--Iron & 45$^{\circ}$ &  26$^{\circ}$ & +/- 30$^{\circ}$ (15$^{\circ}$ to 75$^{\circ}$)  \\
Iron--Copper& 85$^{\circ}$ &  71$^{\circ}$ & +/- 30$^{\circ}$ (45$^{\circ}$ to 90$^{\circ}$)  \\
Copper--Copper & 75$^{\circ}$ &  60$^{\circ}$ & +/- 30$^{\circ}$ (45$^{\circ}$ to 90$^{\circ}$)  \\
Iron--Iron & 75$^{\circ}$ &  60$^{\circ}$ & +/- 30$^{\circ}$ (45$^{\circ}$ to 90$^{\circ}$)  \\
\end{tabular}
\end{ruledtabular}
\label{resultsvtangentlaw}
\end{table}

\section{Making Predictions}

Making predictions can provide students with several exercises. Students can use the below derivation to predict the angles for various pairs of conductors. Advanced students can themselves derive the Tangent Law from the Principle of Least Resistance for assigned sizes of conductors.

In optics, to make a prediction for Snell's Law for two media in series, only the indices of refraction and the incident angle are required. However, in the case of thermal refraction, it is not possible \emph{a priori} to establish either of the angles as an independent variable. However, additional information is known in this demonstration: the dimensions of the conductors. So it is possible to predict where the path of maximum heat flow will intersect the boundary of the two thermal conductors, a point we call $x$. (Predicting the equivalent of $x$ for Snell's Law involves solving a quartic equation that is beyond the scope of this paper).

Since the derivation of the Tangent Law of heat refraction is less commonly known than the derivation of Snell's Law, we provide a derivation in part drawn from Tan and Holland\cite{Tan1990}, but using the set-up of our demonstration. Length $L$ is the longest dimension and height $h$ refers to the two identical shorter dimensions (one of which cancels out) of the conductors. We define thermal resistance $R$ as:
	\[
		R = \frac{\rho l}{A},
	\]
where $\rho$ is the thermal resistivity (1/thermal conductivity $k$) of the material, while $l$ is the length and $A$ the cross-sectional area of a ``tube'' of heat flux, respectively.  As discussed above, the greatest amount of heat will flow through a tube along the path of least thermal resistance. We assume that the thermal resistivity of each particular material is constant. Otherwise, if thermal resistance varies continuously throughout the media, we have to use calculation of variations techniques, just as would be the case when light propagates in non-uniform media.\cite{Weinstock1974}

We have two media with resistivities $\rho_1$ and $\rho_2$. Assume that heat flux travels between two fixed points $P$ and $Q$ located in the two media.  Let $A$ be the slant area of the flux tubes intercepted by the interface.  It follows then that the cross-sectional areas of the tubes of flux in the two media are:
	\[
		\begin{aligned}
			A_1 &= A \cos \theta_1 \quad \mbox{and}\\
			A_2 &= A \cos \theta_2.
		\end{aligned}
	\]
As a result, we obtain the thermal resistance as:
	\[
		\begin{aligned}
			 R 	&= \frac{\rho_1 l_1}{A_1} + \frac{\rho_2 l_2}{A_2} \\
			 	&= \frac{\rho_1 l_1}{A \cos \theta_1} + \frac{\rho_2 l_2}{A \cos \theta_2}.
		 \end{aligned}
	\]
Since 
	\[
		\begin{aligned}
			\cos \theta_1 &= \frac{h_1}{l_1} \quad \mbox{and}\\
			\cos \theta_2 &= \frac{h_2}{l_2},
		 \end{aligned}
	\]
we obtain
	\[
		\begin{aligned}
			 R 	&= \frac{\rho_1 \,l_1}{A \,\,\frac{h_1}{l_1}} + \frac{\rho_2\,l_2}{A \,\, \frac{h_2}{l_2}} \\
			 	&= \frac{\rho_1\, l^2_1}{A \, h_1} + \frac{\rho_2 \,l^2_2}{A \,h_2}.
		 \end{aligned}
	\]
Then substituting  
	\[
		\begin{aligned}
			l_1 &= \sqrt{x^2 + h^2_1} \quad \mbox{and} \\
			l_2 &= \sqrt{(L-x)^2 + h^2_2} , \\
		 \end{aligned}
	\]
into the equation, we obtain
	\begin{equation}\label{E:thermal}
		\begin{aligned}
			 R 	&= \frac{\rho_1\, (x^2 + h^2_1)}{A \, h_1} + \frac{\rho_2 \,[(L-x)^2 + h^2_2]}{A \,h_2}.
		 \end{aligned}
	\end{equation}
Differentiating~\eqref{E:thermal} with respect to $x$ once, we obtain
	\begin{equation}\label{E:first_derivative}
		\frac{dR}{dx} = \frac{2 \rho_1 \,x}{A h_1} - \frac{2 \rho_2 \,(L-x)}{A h_2}.
	\end{equation}
Differentiating again, we obtain:
	\[
		\frac{d^2 R}{dx^2} = \frac{2 \rho_1}{A h_1} + \frac{2 \rho_2}{A h_2},
	\]
which is positive.  So $R$ achieves a minimum at $x$ where $x$ satisfies
	\[
		\frac{dR}{dx} = 0.
	\]
So setting~\eqref{E:first_derivative} to $0$, we get:

	\begin{equation}\label{E:pre_x}
			\frac{\rho_1 \, x}{h_1} = \frac{\rho_2 \, (L-x) }{h_2} 
	\end{equation}
and therefore,  when substituting $1/k$ for $\rho$, is equivalent to the Tangent Law~\eqref{E:tangent_law_w_k} shown earlier:

	\begin{equation}\label{E:tangent_law}
		\rho_1 \tan \theta_1 = \rho_2 \tan \theta_2.
	\end{equation}
However, we wish to predict the angles, which means we will back up a step and solve for $x$. We rearrange~\eqref{E:pre_x}, noting for our demonstration that $h_1$ = $h_2$ (hereinafter simply $h$) and thus cancels out:
	\[
			x=  \left(\frac{\rho_2}{\rho_1}\right)  (L-x) 
	\]	
so that
	\[
			x  = \frac{L}{ \left(\frac{\rho_1}{\rho_2} + 1\right)}. 
	\]	
Then the angles will be:
	\begin{equation}\label{E:theta_1}
			\theta_1  = \arctan{\frac{x}{h}},
	\end{equation}
	
	\begin{equation}\label{E:theta_2}
			\theta_2  = \arctan{\frac{L - x}{h}}.
	\end{equation}	
We now have enough information to predict the angles for a series of two conductors.

\section{Simulation}

Two versions of a simulator for the Tangent Law have been developed, one in Ruby for those who wish to examine or modify the code, and the other in Ruby on Rails for those who wish to explore, but not code. The simulations take as parameters the conductor's material, length, width (and implicitly height). Suitable default values are provided and the simulations look up the coefficient of conduction for each material from internal values.

\subsection{Ruby simulation allows students to ``hack'' the code}

The version of the simulation written in Ruby is suitable for students who are willing to learn or know how to code. Parameters, such as material or conductor length, can be adjusted by manually altering the code. Students can add additional factors or try to create their own hypothesis for comparison with experimental data. Below is a snippet of Ruby code from the simulation:

\begin{verbatim}
# Run the simulation. 
  rho1 = 1.0/thermalconductivity1  # thermal resistance of conductor 1
  rho2 = 1.0/thermalconductivity2  # thermal resistance of conductor 2
  x  = length / ((rho1/rho2) + 1.0)
  theta1 = atan(x/height)
  theta2 = atan((length - x)/height)
\end{verbatim}

The Ruby language simulation is written in version 1.9.2. It has been placed on Github in a publicly available, open-source repository. The code can be downloaded and run on the user's machine. A command line or terminal utility is required, and Ruby 1.9.2 or higher will need to be installed if it is not already present. The Ruby code can be altered or branched as desired. The Ruby simulation code can be found at: \url{<https://github.com/mciotola/conductors_in_series_analytical>}

\subsection{Ruby on Rails simulation}

Another version of the simulation has been written in Ruby on Rails, which is ready for immediate use on the world wide web. It can be used by anyone and works with most web browsers. Neither a command line nor knowledge of any code is required. It  is useful for anyone who wants to quickly make predictions for the above experiment using the Tangent Law, but is especially suitable for students who do not know how to run programs.

The Ruby on Rails simulation is written in version 3.1 (using Ruby version 1.9.2) and placed on the internet as a web application. Its Start view is shown in Fig.~\ref{simulation_view_start}. The Ruby on Rails simulation can be found at: \url{<http://www.heatsuite.com/?page_id=118>} 

\begin{figure}[h!]
\centering
\includegraphics[width=5in]{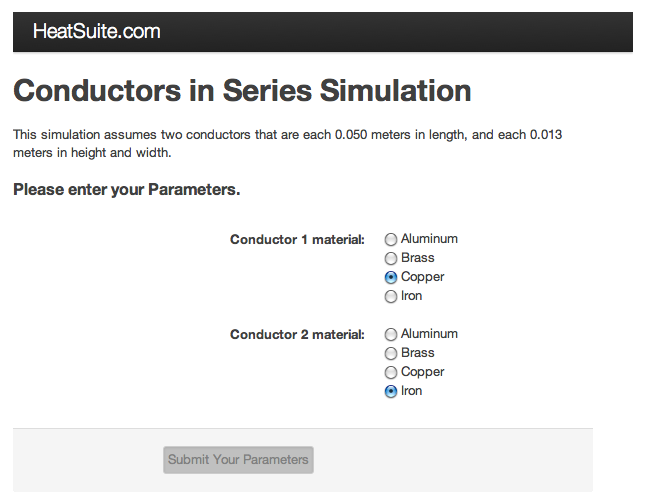}
\caption{Ruby on Rails simulation Start view}
\label{simulation_view_start}
\end{figure}

Experimental results are tested against the Tangent Law simulation in Table~\ref{simtangentlaw1} and Table~\ref{simtangentlaw2}.  Note that each expected angle is generated solely from the Tangent Law independently of the other angle, so only the uncertainty of that angle is applicable. There is an uncertainty associated with the dimensions of the conductors, but it is modest compared to that of the angle measurements. This allows us to reduce the uncertainty to $\pm 15^{\circ}$.
\begin{table}[h!]
\centering
\caption{Comparison  of observed results for Angle $\theta_1$ with simulated values}
\begin{ruledtabular}
\begin{tabular}{l c c c}
Pair of materials & Observed Angle $\theta_1$ & Simulated Angle $\theta_1$ & Total Uncertainty (Range)\\
\hline
Copper--Iron & 75$^{\circ}$ &  73$^{\circ}$ & +/- 15$^{\circ}$ (60$^{\circ}$ to 90$^{\circ}$) \\
Iron--Copper& 30$^{\circ}$ &  33$^{\circ}$ & +/- 15$^{\circ}$ (15$^{\circ}$ to 45$^{\circ}$)\\
Copper--Copper & 60$^{\circ}$ &  63$^{\circ}$ & +/- 15$^{\circ}$ (45$^{\circ}$ to 75$^{\circ}$) \\
Iron--Iron & 60$^{\circ}$ &  63$^{\circ}$ & +/- 15$^{\circ}$ (45$^{\circ}$ to 75$^{\circ}$)\\
\end{tabular}
\end{ruledtabular}
\label{simtangentlaw1}
\end{table}

\begin{table}[h!]
\centering
\caption{Comparison  of observed results for Angle $\theta_2$ with simulated values}
\begin{ruledtabular}
\begin{tabular}{l c c c}
Pair of materials & Observed Angle $\theta_2$ & Simulated Angle $\theta_2$ & Total uncertainty (range) \\
\hline
Copper--Iron & 45$^{\circ}$ &  33$^{\circ}$ & +/- 15$^{\circ}$ (30$^{\circ}$ to 60$^{\circ}$) \\
Iron--Copper& 85$^{\circ}$ &  73$^{\circ}$ & +/- 15$^{\circ}$ (70$^{\circ}$ to 90$^{\circ}$) \\
Copper--Copper & 75$^{\circ}$ &  63$^{\circ}$ & +/- 15$^{\circ}$ (60$^{\circ}$ to 90$^{\circ}$) \\
Iron--Iron & 75$^{\circ}$ &  63$^{\circ}$ & +/- 15$^{\circ}$ (60$^{\circ}$ to 90$^{\circ}$)\\
\end{tabular}
\end{ruledtabular}
\label{simtangentlaw2}
\end{table}

\begin{figure}[h!]
\centering
\includegraphics{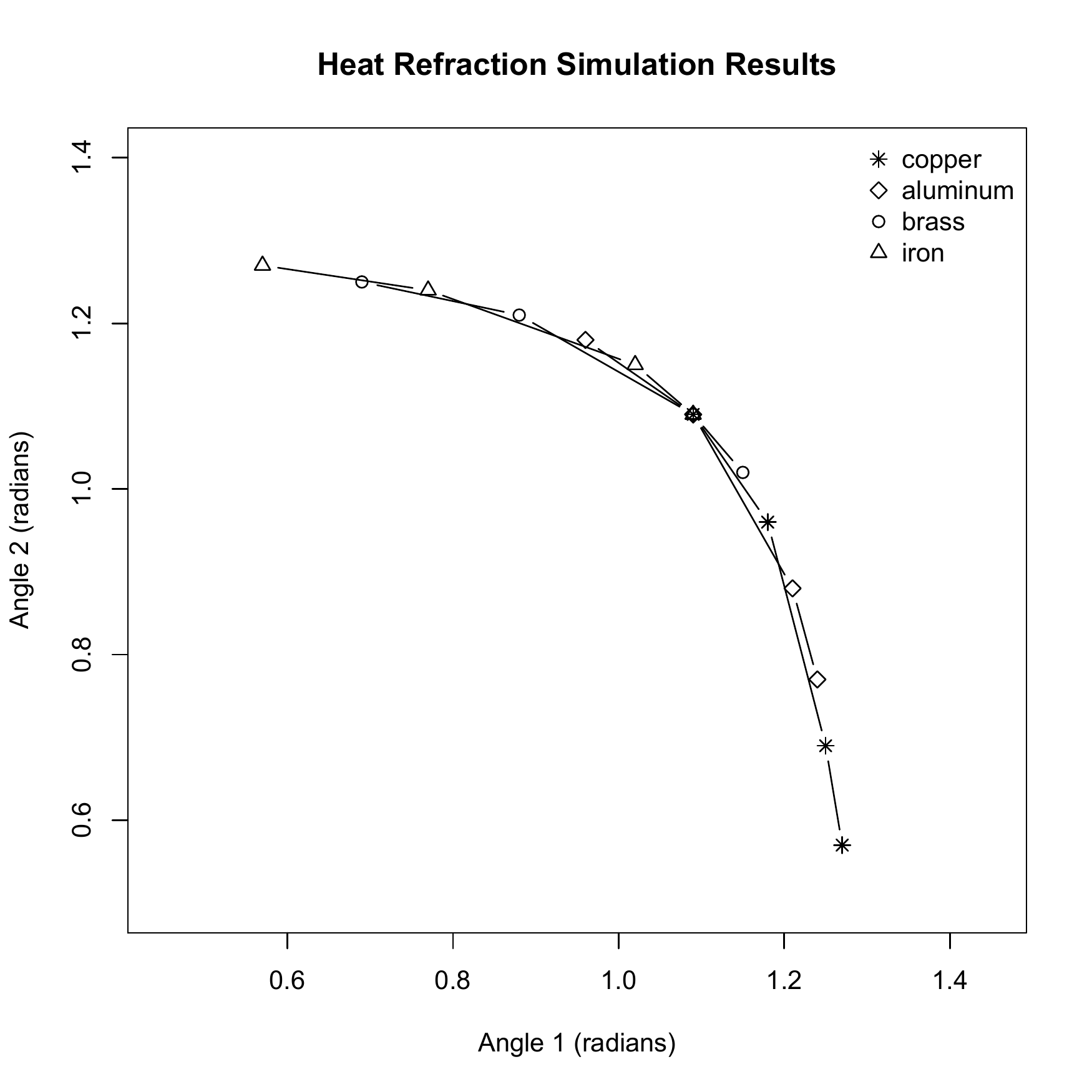}
\caption{Plot of Angle 2 versus Angle 1 for Pairs of Conductors}
\label{thermal_plot_01}
\end{figure}

Results using this simulator can be found for copper paired with various other conductors in Table~\ref{copper}. A plot of simulator results shown in Fig.~\ref{thermal_plot_01} relates the values for Angles $\theta_1$ and $\theta_2$. Each data point shows the value of Angles $\theta_1$ and $\theta_2$ for a particular pair of conductors. Each line connects a series of pairs, where each series contains the same conductor material for at least one pair member. The common material is shown by symbol, while the other conductors are arranged by decreasing conductivity, from left to right. 

Pairs that include lower conductivity iron dominate the upper left part of the curve while pairs containing higher conductivity copper dominate the lower right part. This is consistent with a greater angle indicating a longer route, in turn indicating greater conductivity. All of the materials share a common point at (1.09 rad, 1.09 rad), where both conductors in a pair comprise the same material. The location of this common point is purely a function of the dimensions of the conductors, not their materials.

\begin{table}[h!]
\centering
\caption{Simulated results for material pairs containing copper}
\begin{ruledtabular}
\begin{tabular}{l c c c c}
Materials & Angle $\theta_1$ (radians) & Angle $\theta_2$ (radians) & Angle $\theta_1$ & Angle $\theta_2$ \\
\hline
Copper--Copper & 1.09&  1.09  & 63$^{\circ}$ & 63$^{\circ}$  \\
Copper--Aluminum & 1.18 &  0.96  & 67$^{\circ}$ & 55$^{\circ}$  \\
Copper--Brass & 1.25 &  0.69  & 72$^{\circ}$ & 40$^{\circ}$  \\
Copper--Iron & 1.27 &  0.57  & 73$^{\circ}$ & 33$^{\circ}$  \\
\end{tabular}
\end{ruledtabular}
\label{copper}
\end{table}

\section{Conclusions and Educational Applications}

This colorful classroom experiment demonstrates variational principles to introductory physics students, the wave-particle duality of matter, and an important characteristic of heat flow. It is simple to set up and uses easily-obtained materials. If performed in conjunction with a classic Snell's Law demonstration, it also illustrates how disparate areas physics such as optics and thermodynamics are related by overarching variational principles such as the Principle of Least Time.

Further exercises are possible to analyze the results. For example, an estimate of total heat flow can be made by attempting to divide the flow into several paths, and then using Fourier's Law to calculate relative heat flows through each path. Another example is to determine how much heat flow would decrease if most (or all) of the heat followed an alternative path than that indicated by the Principle of Least Resistance. Further questions can also be posed. For example, how can the results be explained in terms of the discussion in Feynman's Lecture on the Principle of Least Action? In what other situations might refraction be observed? Possible extensions include using a series of more than two conductors, or conductors with differing dimensions.

\appendix*   

\section{Why Ruby?}

\subsection{Ruby}

Ruby is a good first language to learn, because it is simple to read, understand and use.  It is quantitatively robust, and it contains many of features of modern, higher level languages such as object-oriented programming. Computer scientist Yukihiro Matsumoto developed Ruby and first released it in 1995. Ruby draws from Perl, Smalltalk, Ada, and Lisp. \cite{Ruby-Lang} When creating Ruby, Matsumoto strived to develop ``a scripting language that was more powerful than Perl, and more object-oriented than Python ... Ruby is designed to be human-oriented. It reduces the burden of programming. It tries to push jobs back to machines. You can accomplish more tasks with less work, in smaller yet readable code.''\cite{Steward} The clean, plain-English code of the Ruby language makes learning its basics easy and intuitive. Beginners can start with a free, 20 minute online course at the Ruby Lang site:  \url{<http://www.ruby-lang.org/>}.\cite{Ruby-Lang} 

\subsection{Ruby on Rails}

Ruby on Rails\cite{ROR} is an Model-View-Controller (MVC) framework created in 2013, and allows for the rapid development of sophisticated web applications. There are tens of thousands of live Ruby on Rails applications. It is open source, and reputable sites offer free web hosting. Instructors can easily develop their own custom applications and educational tools.

\begin{acknowledgments}
We gratefully acknowledge Prof. James Lockhart at San Francisco State University who supervised the initial literature search. We also thank Prof. Roger Bland at San Francisco State University who suggested the title for this paper. We also appreciate the special efforts of the library staff at the NASA Ames Research Center to preserve their collection of vital works in thermodynamics and heat transfer and thank them for access to those works.
\end{acknowledgments}

\end{document}